\documentclass[aps,prl,reprint,groupedaddress]{revtex4-2}
\input epsf.sty
\usepackage{graphicx,amsmath,amssymb,xcolor,hyperref,ulem}
\hypersetup{colorlinks=true,linkcolor=blue}
\definecolor{red}{cmyk}{0,0.8,1,0}
\definecolor{blue}{cmyk}{1,0.5,0,0}
\definecolor{green}{cmyk}{0.97,0,0.75,0}
\definecolor{cyan}{cmyk}{0.8,0,0,0}
\definecolor{magenta}{cmyk}{0.1,0.7,0,0}
\definecolor{yellow}{cmyk}{0.1,0.05,0.9,0}
\definecolor{orange}{cmyk}{0,0.5,1,0}
 
\def\lromn#1{\uppercase\expandafter{\romannumeral#1}}

\begin{document}

%{\red Red for modified}

%{\blue Blue for modified this time}

%\preprint{magnetization-renp}

\title{
Parametrically amplified super-radiance towards hot big bang universe\\
}



\author{Motohiko Yoshimura}
\email{yoshim@okayama-u.ac.jp}
\affiliation{Research Institute for Interdisciplinary Science,
Okayama University \\
Tsushima-naka 3-1-1 Kita-ku Okayama
700-8530 Japan}

\author{Kunio Kaneta}
\email{kaneta@ed.niigata-u.ac.jp}
\affiliation{
Faculty of Education, Niigata University, \\
Niigata 050-2181, Japan}

\author{Kin-ya Oda}
\email{odakin@lab.twcu.ac.jp}
\affiliation{Department of Mathematics, 
Tokyo Woman's Christian University, \\
Tokyo 167-8585, Japan}


\date{\today}

\begin{abstract}
We propose a mechanism of preheating stage after inflation,
 using a new idea of parametrically amplified super-radiance.
Highly coherent state, characterized by macro-coherence of scalar field coupled to 
produced massless particle in pairs, is created  by parametric resonance effects
associated with field oscillation around its potential minimum,
within a Hubble volume.
The state is described effectively by the simple Dicke-type of super-radiance model,
and super-radiant pulse is emitted within a Hubble time, justifying neglect
of cosmic expansion.
Produced particles are shown to interact to change  their energy and momentum distribution
 to realize thermal hot big bang universe.
A long standing problem of heating after inflation may thus be solved.
A new dark matter candidate produced at the emergence of thermalized universe is 
suggested as well.

\end{abstract}

%\keywords{}

\maketitle


%\subsection{}
%\subsubsection{}

%{\bf Introduction} \hspace{0.3cm}

%\footnote{Key words }

\setcounter{footnote}{0}

\vspace{0.2cm}
{\bf Introduction} 
\hspace{0.2cm}
Inflationary universe scenario solves outstanding problems that
face the beginning of hot big bang, notably flatness and horizon \cite{inflation}
\cite{inflation textbook}.
At the end of inflation there is essentially nothing left except a causally
connected large space-time.
The inflation thus requires further a mechanism that connects it to hot thermal universe
dominated by essentially massless particles,
fostering important events of generating the baryon asymmetry and nucleo-synthesis.

A standard mechanism towards hot big bang after inflation is 
copious  particle production due to
parametric resonances towards the end of oscillatory phase of
scalar field that governs inflation.
Despite of many interesting theoretical attempts and numerical simulations,
\cite{dk},  \cite{bran}, \cite{kls}, \cite{my 95 and fkyy}, \cite{bh}, \cite{tkachev},
\cite{preheating review},
there is no convincing picture of how a hot big bang is realized after inflation 
\cite{effect of hubble expansion}.

Our present work clarifies the presence of highly coherent state 
characterized by macro-coherence 
\cite{macro-coherence}. This  coherence is  realized by
induced polarization of  massless particle pairs due to
parametric resonance,
and super-radiant  emission of massless particle pairs
releases  nearly all stored energy at an instant after macro-coherence
 is fully developed.
The macro-coherence works in the entire Hubble volume, hence much
more efficient than the usual Dicke super-radiance limited by wavelength of emitted photon
\cite{dicke}, \cite{gross-haroche}, \cite{sr textbook}.
Momentum distribution of produced massless particles
at this stage of preheating is far away from
thermal distribution, and re-distribution of momenta is 
shown to occur by interaction among produced particles, which require conditions
readily realizable.

Throughout this paper we use the unit of $\hbar = c = k_B = 1$.

\vspace{0.2cm}
{\bf Inflaton field and its coupling to matter}
\hspace{0.2cm}

While our results may be applied to a broader class of models, 
we focus on essential features of the present work, and specify 
the Lagrangian of the scalar field $\chi$, called inflaton, 
and its coupling to
the matter Lagrangian, ${\cal L}_m$, of the Standard Model or Grand Unified Theory (necessary if lepto- or
baryo-genesis is to be implemented). In doing so, we resort to the 
principle of conformal coupling of $\chi$ to the trace of the energy-momentum tensor determined by
${\cal L}_m$.
Its simplest realization is provided by Jordan-Brans-Dicke gravity \cite{jbd} of the Lagrangian form,
$- \sqrt{g_J} M_{\rm P}^2 F(\chi)\, R_J/2$, 
(with $M_{\rm P}$ the reduced Planck mass),
transformed by Weyl rescaling,
$ g_{J\mu\nu} = g_{\mu\nu} F^{-1}(\chi)$, to the Einstein metric frame.
We further add a potential term for inflaton, extending 
the original theory \cite{jbd}, thus called
extended Jordan-Brans-Dicke gravity (eJBD) \cite{ejbd}.
eJBD field appears naturally  from quantum gravity theory of super-string
in higher dimensional space-time.
Our field-redefined form of Lagrangian in the Einstein metric is thus
\begin{eqnarray}
&&
{\cal L} = \sqrt{-g} \left(- \frac{ M_{\rm P}^2}{2}R+ 
 \frac{1}{2} (\partial \chi)^2 - V_{\rm HO}(\chi)
 \right.
 \nonumber \\ &&
 \left.
+ F^2(\gamma \chi) {\cal L}_m(\varphi, g_{\mu\nu} F^{-1}(\gamma \chi)\,)
\right)
\,,
\end{eqnarray}
with $\varphi$  generically denoting matter fields.

The potential $V_{\rm HO}(\chi)$ is harmonic  oscillation (HO) approximation to a more complicated inflaton potential $V_{\chi}$, on which we elaborate later.
The coupling constant $\gamma$ of mass dimension $-1$ is taken of order $1/M_{\rm P}$, hence scalar-matter field coupling
is of gravitational strength.

Since at early epochs right after inflation all particle masses in the Standard Model 
effectively vanish, direct coupling to 
the matter fields are absent in the trace of energy-momentum tensor
at the classical level. 
There is however quantum correction, known as
the trace anomaly, giving rise to \cite{ps}
\begin{eqnarray}
&&
- \gamma \chi \,T^{\mu}_{\mu}
\,, \hspace{0.3cm}
T^{\mu}_{\mu} = - 
 \frac{b_i \alpha_i}{16\pi} (F^i_{\rho\sigma})^2
\,,
\label {trace anomaly of gauge-pair}
\end{eqnarray}
with $\alpha_i = g_i^2/4\pi$.
Here $ b_i $'s are  coefficients of the renormalization group
$\beta$-functions.
For instance, $b_{\rm em} = 11/3$ for the photon.
Matter fields that enter in this trace anomaly are squared field strength of
electroweak gauge bosons,
QCD gluons in standard model and plus lepton- and baryon-number
violating gauge bosons in grand unified theories,
all denoted by the index $i$. 
We ignored coupling to fermion pairs since parametric resonance effect is 
 negligible due to the Pauli blocking, 
and further omitted possible longitudinal  Higgs boson contribution for simplicity.

Relevant energy scale for the use of the trace anomaly is at the expected reheat
temperature $\sim 10^{16}$GeV, hence the renormalization group analysis
for $g_i^2/4\pi$ is needed.
We shall take the popular unified coupling $ \alpha_{\rm G} \sim 1/35$
in SO(10) grand unified models \cite{so10-1}, \cite{so10-2}).
SO(10) fermion (including right-handed neutral lepton $N_R$) loop contributions give 
 \begin{equation}
\sum_i \frac{b_i \alpha_i}{16\pi} \equiv 
c_t \frac{\alpha_{\rm G}}{16 \pi} \sim - 0.014
\,,
\end{equation}
to be multiplied by a common squared gauge field strength
$F_{\rho \sigma}^2 $.

Great merits of this eJBD approach \cite{koy 2ejbd} are (1) a cosmological constant present in
the Jordan-frame may be suppressed in the Einstein frame by a judicious choice
of $F(\chi)$ \cite{exponential potential}, thus evading its fine-tuning problem,
and (2) extension of two field eJBD theory may lead to simultaneous solution to inflation and accelerating  universe near the present epoch.

The rest of our analysis holds in a more general class of models
that give inflaton coupling to the trace of energy-momentum tensor.

\vspace{0.2cm}
{\bf Macro-coherent state due to parametric amplification}
\hspace{0.2cm}
At the end of inflation, inflaton field $\chi$ starts to oscillate around its potential
minimum which is well described in the harmonic oscillator (HO).
HO approximation gives in the flat FLRW metric a damped oscillation due
to the Hubble friction $\sim 3 H \chi'$ in the $\chi-$ 
evolution equation, which
takes a simple sinusoidal form $\chi_0 \cos (m_{\chi} t )$ within the Hubble time
$1/H$, where $\chi'\equiv d\chi/dt$.
We shall confine, for the moment, to the case within Horizon scales.

Due to the space-translational invariance of background inflaton field, 
the quantum system of gauge bosons may be Fourier-decomposed into momentum $\vec{k}-$modes
and treat each mode independently.

Gauge fields in the radiation gauge of $A_0=0, \vec{\nabla}\cdot \vec{A} =0 $
satisfy 
\begin{eqnarray}
&&
\frac{d^2}{dt^2} \vec{A}_k^i + k^2 \vec{A}_k^i + \frac{K_i \chi'}{1+ K_i \chi}
 \frac{d}{dt} \vec{A}_k^i = 0
\,, 
\\ &&
K_i =  \gamma \frac{b_i\alpha_i}{4\pi}
\,,
\end{eqnarray}
neglecting small tri-linear coupling terms for non-Abelian gauge fields.
This equation is  recast to,  by denoting 
two components  of vector fields, $\vec{D}_k^i = (1+ K_i \chi)^{-1/2}\vec{A}_k^i $
by a single $u_k$, 
\begin{eqnarray}
&&
\frac{d^2}{dt^2} u_k + k^2 u_k - \frac{K_i}{2}  \chi'' u_k = 0
\,,
\label {mathieu eq}
\end{eqnarray}
with the  approximation $\frac{K_i  \chi''}{ 1+ K_i \chi} \approx K_i \chi'' $
and dropping higher order terms $ 3\left( (K_i \chi')/(1 + K_i \chi) \right)^2/4$.
When $\chi(t)$ is of the sinusoidal form in the HO region, 
the differential equation
for $u_k$ has a periodic function $\chi(t)''$ as a coefficient.

The standard form of Mathieu equation \cite{mathieu eq} is defined by scaling 
shifted time  and
dimensional parameters in (\ref {mathieu eq}), and is written as
\begin{eqnarray}
&&
\left( \frac{d^2}{d\tau^2} + (a - 2 q \cos 2 \tau ) \right) u_k = 0
\,, 
\label {mathieu eq 2}
\\ &&
 \tau =  \frac{ m_{\chi}}{2} (t + \pi)
\,, \hspace{0.3cm}
a = \frac{4 k^2} {m_{\chi}^2} 
\,, \hspace{0.3cm}
q=  K_i \chi_0
\,.
\end{eqnarray}
$\chi_0$ is the initial amplitude of sinusoidal function $\chi(\tau) $. 
The  energy $k$
of massless gauge bosons is equal to the half of the parent $\chi-$particle at rest.
This gives the parameter in the Mathieu equation; $a=1$.

Solutions of Mathieu equation (\ref{mathieu eq 2}) are characterized in terms of
band structure in the $(q,a)$ plane where
bounded Bloch-type solutions (important in condensed matter physics
when time is replaced by spatial coordinate) and exponentially unstable
solutions alternate with their boundaries given by $a(q) $.
The exponential growth is called parametric amplification in our terminology,
also called broad resonance in other literatures.
The small amplitude region of $q\rightarrow 0$ gives discrete unstable  narrow bands
around $a \sim n^2\,, n=0,1,2, \cdots$, and
 is interpreted as a collapsed n-body $\chi$ decay;
 $n \chi \rightarrow A_k + A_{-k} $ \cite{my pr-perturbation}.
On the other hand, the largest unstable bands are in the large $q-$amplitude
with $ a \leq O(2) q$.
The band structure and detailed behaviors of solutions are
semi-analytically or numerically analyzed as in \cite{mathieu eq}.
Mathematical softwares are available for unstable Mathieu solutions as well.

The phase coherence of produced gauge fields
is best studied in the Schroedinger picture of quantum gauge system
introduced in \cite{my 95}.
We use wave functional of each momentum-mode $\Psi (q_k; t)$
that satisfies
\begin{eqnarray}
&&
i \frac{\partial}{\partial t} \Psi (q_k;  t) = - \frac{1}{2} \frac{\partial^2}{\partial^2 q_k}
\Psi (q_k;  t) + \frac{ k^2 }{2}  q_k^2  \Psi (q_k;  t) 
\end{eqnarray}
The total wave functional is the direct product of each momentum-mode.
Variable $q_k$  corresponds to momentum decomposed inflaton $\chi$ field
and may be regarded as HO coordinate.

The gaussian ansatz for variable $q_k$  of the wave functional 
leads to
\begin{eqnarray}
&&
\Psi (q_k; \tau) = \frac{1}{|u_k|} \, \exp[- \frac{q_k^2}{2 |u_k|^2} \left(
\pi - \frac{i}{2} \frac{d}{d\tau } |u_k|^2 \right) ]
\end{eqnarray}
where $u_k$ is 
 proved to satisfy the Mathieu equation of type (\ref{mathieu eq 2})
written in terms of rescaled time $\tau= (m_{\chi} t + \pi)/2$.
This is exact result when the Hubble friction can be neglected in
the HO region.
When the parameter $(a,q)$ of solution 
$u_k$ belongs to a instability band, magnitudes of the gaussian  width 
$\sim |u_k|^2$ of the
wave functional $| \Psi (q_k; \tau )|^2$ exponentially grows in time,
indicating copious gauge boson production.

A striking feature of wave functional is a constant phase 
given by $ d \ln |u_k|^2/d\tau /2$
when the Mathieu solution is 
purely exponentially increasing within instability bands.
The phase thus determined may differ in different momentum-mode of $\vec{k}$,
but the Mathieu chart has degeneracy in direction of vector $\vec{k}$,
hence modes of the same vector magnitude have a common phase.

The exponential growth does not continue indefinitely due to
dissipative decay of parent field $\chi$. 
Let us discuss the back reaction of particle pair production against the $\chi$ system.
The effective lagrangian ${\cal L}_{\rm parametric} $ 
should  contain dissipation term due to 
parametrically  amplified two-particle decay and quantum fluctuation $\langle A_k^2 \rangle $ 
necessarily related to the dissipation due to
the fluctuation-dissipation theorem of statistical mechanics.
The effective lagrangian  generally takes the form,
\begin{eqnarray}
&&
{\cal L}_{\rm parametric} =
\int \frac{d^3 k}{(2\pi)^3} 
\nonumber \\ &&
\left( 
\frac{1}{2} \Gamma_k^{\rm R} (\chi) \chi \chi'
%\right. \nonumber \\ && \left.
+
c_t \frac{\alpha_{\rm G}}{16 \pi} 
\, \gamma m_{\chi}^2\, \langle A_k^2 \rangle (t) \chi 
\right)
\,.
\label {parametric effect}
\end{eqnarray}
Quantum fluctuation is typically 
related to Rayleigh dissipation $\Gamma_k^{\rm R} (\chi) $ \cite{goldstein}
by $ \langle A_k^2 \rangle (t) \propto e^{\Gamma_k^R t}$.

Lagrangian of Rayleigh dissipation
$\propto  \chi \chi'$ is necessarily time-reversal violating, while
quantum fluctuation 
generates  a kind of induced polarization $\propto \langle A_k^2 \rangle$ 
in medium within the  system initially dominated by $\chi$ field alone.
In this  $\sim \chi A_k^2$ interaction model
the induced polarization acts as a time dependent external force to 
the $\chi$ system.
The coherence of produced gauge bosons
 for definite momentum-pair mode $(\vec{k}, - \vec{k})$ is maintained during the parametric amplification  within instability bands.

 Rayleigh dissipation rate $\Gamma_k^R$ is given by
the decay rate in perturbation theory times 
the enhancement factor $\Delta k/m_{\chi}$, 
which is calculated to give
 \begin{eqnarray}
&& 
\Gamma_k^R = \Gamma_B \frac{\Delta k}{m_{\chi}}
\,, \hspace{0.3cm}
\Gamma_B = (c_t \frac{\alpha_{\rm G}}{16 \pi} )^2
\, \frac{\gamma^2 m_{\chi}^3}{2}
\,,
\label {rayleigh dissipation rate}
\end{eqnarray}
with $\Delta \chi \gg m_{\chi}$ the field dependent width
of relevant instability band,
which may be  numerically  estimated.

We note on how parametric amplification may affect the inflationary stage.
A large exponential amplification and copious particle production
require a sizable oscillation amplitude in HO region.
Outside HO region the potential form may differ.
We shall take a single field version of potential advocated
in \cite{koy 2ejbd}, which reads as
\begin{eqnarray}
&&
V_{\chi} = V_0 \,(\lambda_0 + c \chi^2  )\,e^{ - \gamma \chi}
\,, \hspace{0.3cm}
\lambda_0 >0
\,, \hspace{0.3cm}
c>0
\,.
\end{eqnarray}
Inflation in this potential occurs slightly below a potential maximum
 around $2/\gamma$ (assuming a small $\lambda_0$), but
much above the HO region around $\lambda_0 \gamma/2c$.
Hence,
in the early phase of inflation there is no substantial effect
of parametric amplification.
It is true however that chaotic type  inflation of quadratic potential
is much affected by parametrically amplified super-radiance.

\vspace{0.2cm}
{\bf Application of Dicke super-radiance model}
\hspace{0.2cm}
In order to make relation of parametric amplification to super-radiance  clearer,
we extend the density matrix extending given in \cite{my 95}.
The bilinear projection operator $|\Psi(q_k; \tau) \rangle \langle \Psi(q_k; \tau) |$
onto pure quantum states of \cite{my 95}
is first expanded in terms of most convenient basis, for which we take
the HO energy eigen-states $|n\rangle \sim
(a^{\dagger})^n\, |0 \rangle\,, \langle q_k |0 \rangle  \propto e^{- m_{\chi} q_k^2/2 }$.
$a, a^{\dagger}$ are annihilation and creation operators of HO and
$|0\rangle $ is the ground state of zero-point energy.
The off-diagonal matrix elements, 
$\langle n |\Psi(q_k; \tau) \rangle \langle \Psi(q_k; \tau) |n'\rangle\,, n \neq n'$,
rapidly oscillates in time, and its short time average nearly vanishes.
Thus, a classical system of finite non-vanishing entropy emerges.

The total number of $\chi$ particles within the Hubble horizon is
\begin{eqnarray}
&&
{\cal N}_{\chi} = H^{-3} n_{\chi} \sim H^{-3} \frac{ \rho_{\chi} }{m_{\chi} }
\,,
\label {particle n within h}
\end{eqnarray}
with $n_{\chi}, \rho_{\chi} $ the number and the energy densities of $\chi$ particles.
In HO region the non-relativistic relation $\rho_{\chi} = n_{\chi} m_{\chi}$ is valid.

The state  after parametric resonance effects are fully developed 
 is well described by Dicke model based on the algebra of
angular momentum \cite{dicke},  \cite{gross-haroche}, \cite{sr textbook}.
There are however crucially important differences:
(1) There are only two states in the Dicke model, linear combinations
of the ground and some excited state, while ours has states
of order $M_{\rm P}/m_{\chi}$, typically as large as $O(1 \sim 10^5)$,
(2) inflaton decay products in our model are correlated
 two massless particle pairs unlike a single photon
 in the Dicke case.

In the original Dicke model of super-radiance it is supposed that
laser irradiation generates a coherent state within the light wavelength region,
the region limitation caused by a common
 phase factor of incident radiation $e^{i \vec{k}\cdot \vec{x}}$ with $k \sim 1/{\rm wavelength}$.
Our case of pair production
 has the vanishing phase $e^{i (\vec{k}- \vec{k})\cdot \vec{x}} = 1$ for
relevant back to back pair emission.

The enhanced coherence without the wavelength limitation
 was theoretically suggested and
is called macro-coherence \cite{macro-coherence}. The macro-coherent two-photon
emission  was experimentally verified in \cite{h2 okayama exp}
 for para-H$_2$ molecular transition,
 with rate enhancement of order $ 10^{15}$.

Despite of these differences one can take over nice features
of the Dicke model provided that the wavelength limitation
of coherent region is lifted.

\vspace{0.2cm}
{\bf Phase development and enhanced super-radiant decay rate}
\hspace{0.2cm}
Following Dicke \cite{dicke}, one can introduce the algebra of 
the total angular momentum satisfying 
\begin{eqnarray}
&&
[R_i\,, R_j]= i \epsilon_{ijk} R_k
\,.
\end{eqnarray}
The value $J$ of the total angular momentum 
 is related to product ${\cal N}$ of the $\chi-$particle number
and the number of levels $M_{\rm P}/m_{\chi} $; ${\cal N} = 2 J +1 \approx 2 J $.
This product is further related to the $\chi$ energy  within the Horizon volume, 
hence
${\cal N} m_{\chi} = H^{-3} \rho_{\chi}$. By using (\ref{particle n within h}),
we have
\begin{eqnarray}
&&
{\cal N} = 
 {\cal N}_{\chi}\,\frac{M_{\rm P} }{ m_{\chi}}
\,.
\label {number of inflaton}
\end{eqnarray}
Projection of angular momentum onto an axis has component values $M$
ranging from $M=J $ to $M=-J$, hence the entire range 
being $\Delta M = 2 J + 1 \approx 2J$.
Despite that more than two levels are introduced in our case
 excitation and de-excitation occur between
two energetically adjacent levels, hence $\Delta M = \pm 1$.
Hence one can use the lowering $R_-$ operator for pair emission and
the raising $R_+$ operator for pair excitation between $\Delta M = \pm 1$.
This makes our case and the Dicke case essentially equivalent.

Different momentum $\vec{k}-$modes evolve independently
and one must sum over these contributions in the end.
But if a particular momentum mode evolution is predominantly fast,
this mode alone may take over the entire evolution to
dominantly realize hot thermal universe at the next stage.

The population probability denoted below by $P_M(t)$ of the state $|J, M \rangle $ 
follows the evolution equation
\begin{eqnarray}
&&
\frac{d P_M}{dt}  = \Gamma_k^R \left( |\langle J, M+1 | R^+ | J, M \rangle |^2
P_{M+1}
\right.
\nonumber \\ &&
\left.
 -   |\langle J, M-1 | R^- | J, M \rangle |^2 P_{M-1} \right) 
\,.
\label {rate eq}
\end{eqnarray}
supplemented by well-known relations,
$ |\langle J, M+1 | R^+ | J, M \rangle |^2 = (\frac{N}{2} - M) ( \frac{N}{2} + M +1)$ etc.
One can prove the law of probability conservation,
\begin{eqnarray}
&&
\frac{d}{dt}\, \sum_{M= -J}^{J} P_M = 0
\,.
\end{eqnarray}
This is a coupled set of linear differential equations of a gigantic size
$2J = {\cal N}_{\chi}$ (the quantity given by (\ref{number of inflaton})\,).
After the end of inflation, ${\cal N}_{\chi}$ number of $\chi$ particles
evolve independently within the horizon, 
in particular they may decay at different times.
One needs to match the stage of parametric amplification to
the super-radiant stage smoothly. 

%{number of inflaton}

One may assign HO state $|n \rangle $
of gauge bosons to each $\chi$ particle for this purpose.
Due to energy exchange of $\chi$ and gauge boson, one
can derive the relation $n = J-M$.
The initial state in the Dicke picture may be identified
as distributed $M-$states by the relative weight of  density matrix elements
$\propto \rho_k \left( M,M ; \tau= 0 \right) $,
\begin{eqnarray}
&&
|J, M \rangle_i = \left( \sum_{M'} 
( \rho_k ( M',M' ; 0 ) \right)^{-1/2} 
\nonumber \\ 
&&
\hspace*{0.5cm}
\sqrt{
\rho_k \left( M,M ; 0 \right)}\, | J, M \rangle
\equiv P_M(0) | J, M \rangle
\,.
\label {initial state for dicke}
\end{eqnarray}
When the parametric amplification is fully developed,
density matrix elements $\rho_k \left( M,M ; 0 \right) $ are calculable.

\vspace{1cm}
\begin{figure*}[t]
    \centering
    \includegraphics[width=.4\textwidth]{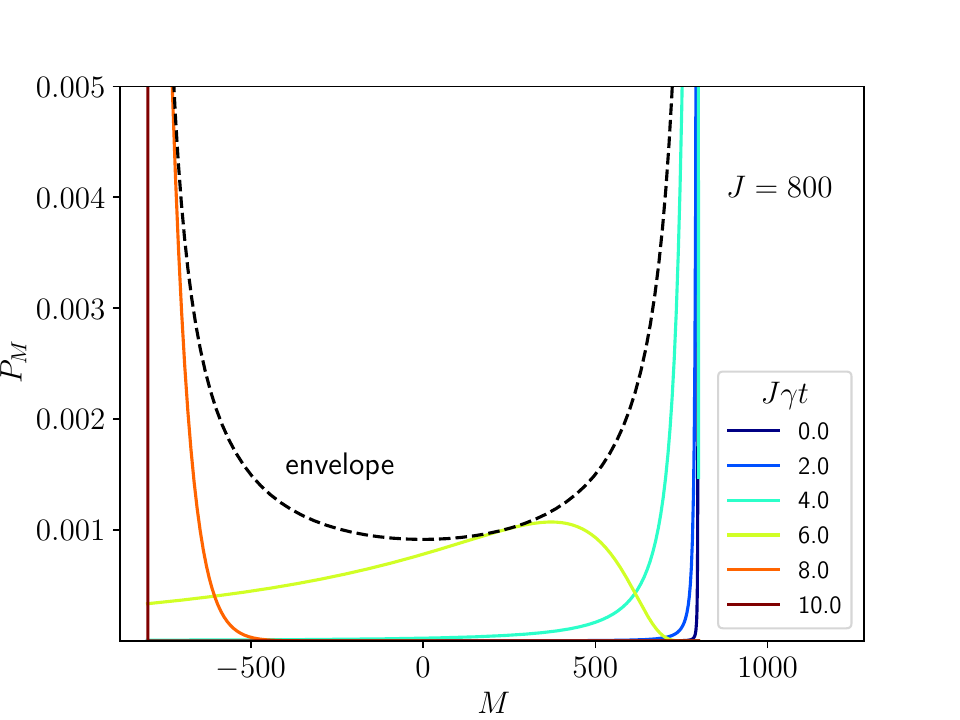}~~~
    \includegraphics[width=.4\textwidth]{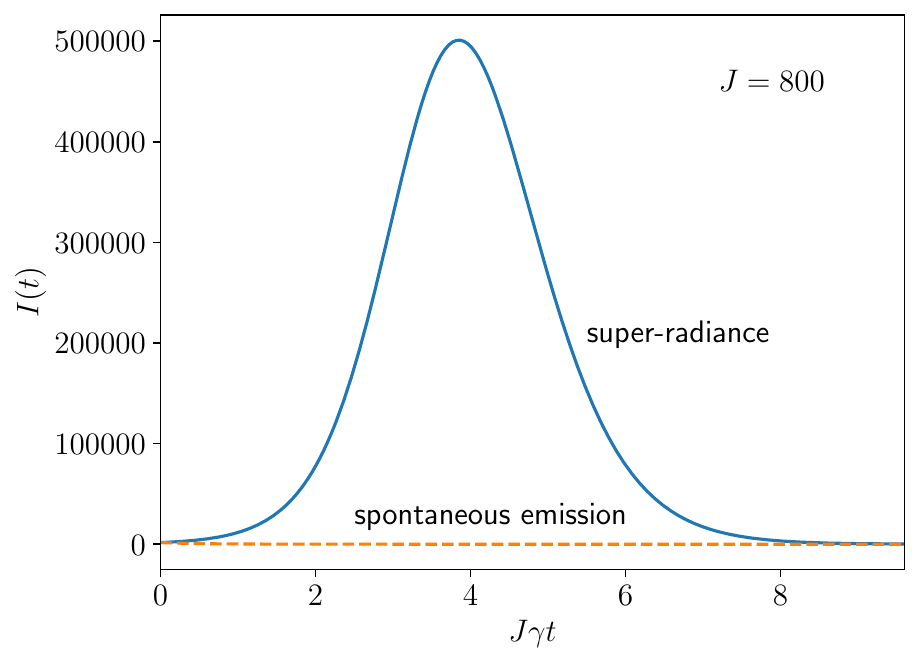}
% \vspace{2.5cm}
     \caption{
     \textbf{(left panel)} Time evolution of $P_M$ as a function of $M$.  
The black dashed line labeled by envelope is the envelope 
of  largest peak values at each $M$.
     \textbf{(right panel)} Time evolution of the radiation intensity,
the one predicted by super-radiance in sold blue and by spontaneous emission
in dashed red.
    }
    \label{fig1}
\end{figure*}

A  systematic method to determine
the weight factors $\rho_k \left(M,M; 0 \right)$ in (\ref{initial state for dicke}) 
shall be worked out in our subsequent longer paper.
Here we are content to concentrate on qualitative features that arise from
much simplified initial state.

Since the fundamental Dicke equation (\ref{rate eq}) is a set of linear differential equations, one can decompose this linear system and later take 
linear combinations of decomposed components.
That this procedure is useful is justified due to a common phase in
$\rho_k \left( M,M ; t_i \right)$ for each $M$.
The simplest initial condition is the Dicke initial condition by
 $P_M(0) = \delta_{M, J}$, a single state
for a specific $J$ value which may be different from the estimated $J= {\cal N}/2$.
This would clarify qualitatively how a collective decay of the $\chi$ system may occur
depending on $J$.

Chain states connected by operation $R^-$ satisfy permutation symmetry
of different $\chi-$particles.
Degeneracy increases each time a chain decay occurs.
This may be interpreted as a further coherence development of the system.

The concept of delayed time in super-radiance 
\cite{gross-haroche}, \cite{sr textbook}, needs to be stressed.
The spontaneous decay governed by the exponential law $e^{- \gamma_B t}$
is characterized by the lifetime of averaged decay time $t_B = 1/\gamma_B$.
In the parametrically amplified decay the perturbative decay rate $\gamma_B$
is replaced by the Rayleigh dissipation rate $\Gamma_k^R$.
On the other hand, the super-radiant decay depletes almost all excited
states at what is called delayed time given by 
$ \ln {\cal N}/{\cal N} \times$ the lifetime, 
which is much shorter
than the lifetime of spontaneous decay for a very large ${\cal N} = 2 J$.
The delayed time $t_d$ is given by
\begin{eqnarray}
&&
t_d = \frac{\ln {\cal N} }{{\cal N} }\, \frac{1}{\Gamma_k^R } 
\,.
\end{eqnarray}
The depletion leaves behind $\chi$ field corresponding to
zero-point oscillation.

A simple estimate of delayed time is possible by taking the Hubble rate
as $H = \Gamma_k^R$ in the formula of ${\cal N}$, (\ref{number of inflaton}),
to give a crudely estimated quantity,
\begin{eqnarray}
&&
{\cal N}
% \sim (\frac{16\pi }{|c_t|\alpha_{\rm G} })^3 
% \frac{\rho_{\chi} }{m_{\chi}^7 \Delta \chi^3 \gamma^6 }
% \sim  1.1 \times 10^{10}
%  (\frac{M_{\rm P}}{m_{\chi}})^7
\sim (\frac{16\pi }{|c_t|\alpha_{\rm G} })^6
\frac{M_P\rho_{\chi} }{m_{\chi}^8 \Delta \chi^3 \gamma^6 }
\sim 4\times 10^7
(\frac{M_{\rm P}}{m_{\chi}})^8
\,,
\label {N estimate}
\end{eqnarray}
highly sensitive to the much unknown $m_{\chi}$ value.
The energy density $\rho_{\chi}$ is taken $(10^{16}{\rm GeV})^4$
from anticipated thermal energy that subsequently follows.
Other adopted parameter values are
 $\gamma = 0.1/M_{\rm P}, \Delta \chi = M_{\rm P}$.
%The super-radiant decay, with the factor $\ln {\cal N}$ included, is thus shortened by more than $O(10^{12})$.

We illustrate in Fig(\ref{fig1})
result of numerical integration of rate equations (\ref{rate eq})
under the Dicke initial condition.
Similar numerical computation has been performed in \cite{occupation number}.
As time proceeds, 
distribution of state changes in time scale of a fraction $O(1/J \gamma)$,
which means after a few successive spontaneous emissions the phase coherence is developed.
At the vicinity of $M\sim 0$, all the states are nearly degenerate in $M$, at which
the pulse-like radiation is emitted as shown in the right panel.
The radiation intensity is estimated by $I(t)=-\frac{d}{dt}\sum_{M=-J}^{M=J}M P_M(t)$,
to give a result in the right panel of Fig(\ref{fig1}).
In this example of a modest value of $J=800$,
the delayed time occurs at $\sim 4/(800 \times$the Rayleigh dissipation rate).
As clear from the right panel of Fig(\ref{fig1}), the macro-coherent
super-radiance terminates copious particle production at times
when spontaneous emission is hardly working.

The condition of neglecting Hubble expansion is justified if the inequality
is satisfied;
$ \Gamma_k^R {\cal N}^2> H $,
incorporating macro-coherent amplified rate $\propto {\cal N}^2$.
At the expected time scale of $H = \Gamma_k^R$,
this condition is equivalent to ${\cal N}^2 > 1$,
which is readily satisfied as indicated in the ${\cal N}$ estimate of
(\ref{N estimate}).

The elegant Dicke's algebraic method is simple and convincing.
Our parametrically amplified super-radiance has an extraordinary merit in
that this simplicity may be applied to the entire universe within the horizon.
We have only sketched how the parametric amplification
prepares macro-coherent super-radiance without
giving details of the initial state for the Dicke algebraic model.
Going beyond this oversimplified approach is numerically demanding,
but is worth of detailed study, for instance for the purpose of
exploring the possibility of primordial black hole formation.

We note a striking difference of parametrically amplified super-radiance
from laser irradiated super-radiance. Our case is a phenomenon of self-organized 
macro-coherence evolution
by inflaton oscillation, unlike the other case of
laser-triggered evolution. It would be of great interest if one can find this kind of
self-organized phenomenon in laboratories.

\vspace{0.2cm}
{\bf Thermalization}
\hspace{0.2cm}
The state of universe within the Horizon volume
right after macro-coherent super-radiance may
be described by massless pairs of many different momenta
$\vec{k}$ with different spread $\Delta^3 k$
(arising from regions deep in instability bands of
the Mathieu chart) at different sites
$\vec{x}$  of spread $\Delta^3 x$.
This is far from the thermal distribution
described by a single temperature.
The universe is however spatially homogeneous, and
particle momenta are  randomly distributed.
Thermalization may occur during the next stage of interaction among
produced particles.
A question now arises whether thermalization is fast enough,
namely, whether thermalization rate is larger than the Hubble rate.

We shall check the consistency condition, $\Gamma > H$, 
taking first an average momentum $T$ and
confirming this to be regarded as a finite temperature.
To estimate thermalization rate $\Gamma(T)$, let us take an example of Compton process
given by total cross section $2\pi \alpha^2 \ln(s/m_e^2)/s$ in high energy limit 
of squared center-of-mass energy $s \gg m_e^2$.
At finite temperatures the estimated cross section is of order, $\pi \alpha^2/T^2$.
Multiplied by the photon number density $n_{\gamma} = 3\zeta(3) T^3/2\pi^2$,
Compton rate at finite temperature is of order, $ 3 \zeta(3) \alpha^2 T/2\pi$.
There are as many as $O(N)$ colliding particles against, say, a fermion, which gives
an effective rate,
\begin{eqnarray}
&&
\Gamma = \frac{3 \zeta(3)}{2\pi}N \overline{\alpha^2} T
\,.
\end{eqnarray}
This should be larger than the Hubble rate,
\begin{eqnarray}
&&
\hspace*{-0.3cm}
H= \frac{\pi }{3}\sqrt{\frac{N }{10 }} \frac{T^2 }{ M_{P}}
\,, \hspace{0.3cm}
M_P =\frac{1}{\sqrt{8\pi G_N}} \sim 2.4 \times 10^{18}\, {\rm GeV}
\,.
\nonumber \\ &&
\end{eqnarray}
The required inequality $\Gamma >  H$ gives a condition,
\begin{eqnarray}
&&
T < \frac{9 \sqrt{10}\, \zeta(3) }{2\pi^2 } \sqrt{N}\, \overline{\alpha^2} M_P
\sim 1.73 \sqrt{N} \overline{\alpha^2} M_P
\,.
\end{eqnarray}
We may take $N=O(100)$ in the Standard Model, $O(500)$ in SO(10) Grand-Unified-Theory models,
and $ \overline{\alpha^2} = O(10^{-4} \sim 10^{-3})$.
The right hand side is expected to be slightly larger than $10^{16}$ GeV.
This shows that thermalized universe of temperature $O(10^{16})$ GeV
may be realized.
It is fortunate that this temperature is sufficiently high such that
subsequent baryo- or lepto-genesis may occur.

\vspace{0.2cm}
{\bf Dark matter candidate}
\hspace{0.2cm}
 As a 
byproduct of the super-radiant decay of $\chi$, a dark matter candidate naturally arises.
Gauge boson pair at a site in our Dicke model forms a triplet
due to the maximal symmetry of the angular momentum state
 \cite{spin triplet}.
Non-trivial topology may be defined 
by mapping the unit sphere in real sphere
onto modulus of the 3-vector of triplet.
It is more likely that topological objects are formed involving
many different sites instead of a single site.
Objects of finite discrete winding number of this mapping may survive
till later epochs of cosmological evolution
despite surface interaction of condensates with surrounding thermal medium.
Remnant objects after recombination become a good candidate of dark matter.
The dark matter thus constructed is made of
$(\vec{k}, - \vec{k})$ modes belonging to stability bands of the Mathieu chart.

During thermal evolution after condensate formation at the end of inflation
accretion of fermions may occur, and it is not clear what sort of
constituents make up dark matter at recombination.
Moreover, their masses may have a wide range.
It is difficult at the moment to determine how to detect these dark matter.
One cannot exclude the possibility that accreting condensates end with
primordial black holes.

Unlike all other dark matter candidates considered in
the literature dark matter objects in this scenario are produced
simultaneous with emergence of the hot universe.
Thus, more detailed investigation may make their 
cosmological relevance more evident.

\vspace{0.2cm}
{\bf Discussion and outlook}
\hspace{0.2cm}
During earlier phases of inflation the harmonic potential approximation
is not valid and analysis based on the Mathieu equation needs to
be modified for discussion in this stage.
Non-harmonic pieces of the original potential, typically 
deviation from the eJBD potential,
$ \chi^2 e^{- \gamma \chi }$,
introduces mode mixing and this may introduce phase de-coherence.
The elegant technique introduced by Dicke is no longer applicable,
and one has to set up the Maxwell-Bloch equation 
 \cite{sr textbook},
\cite{macro-coherence}
to deal with spatial de-coherence.
The mode mixing is however present only in the earlier preheating stage,
and one may forget about this de-phasing in the HO regime.
If the phase coherence is fully developed in HO regime, our approach
presented in this work should be approximately valid.

We assumed that inflation ends with parametrically amplified super-radiance
leaving the scalar field at its potential minimum.
It is not entirely clear whether remnant kinetic field energy is large enough
to override a potential barrier in the eJBD type of potential.
If the barrier crossover occurs, it may give a mechanism of
transforming inflaton to dark energy quintessence field.

 We estimated the reheat temperature and the validity of
various approximations using the model of inflaton  that 
couples to the trace of energy-momentum tensor.
Other models of coupling can be dealt using the same
method as given here, and these give different reheat temperature and
their cosmology may be different.

Investigation of these and many other interesting problems
such as new scenario of lepto-genesis are left to future works.

\vspace{0.5cm}
\begin{acknowledgments}
This research was partially
 supported by Grant-in-Aid  Nos. 21H01107 (KO) and 21K03575(MY)   from the Japanese
 Ministry of Education, Culture, Sports, Science and Technology.

\end{acknowledgments}

%\vspace{0.5cm} {\bf  } \hspace{0.3cm}


\begin{thebibliography}{99}
\bibitem{effect of hubble expansion}
Parametric resonance cannot achieve the radiation-dominated universe by itself since the instability bands shutoff as the universe expands, and thus normally some perturbative 
scalar field interaction is assumed for  decay products to dominate in the
thermal energy of the universe.
\bibitem{exponential potential}
An example for $F(\chi)$ is an exponentially decreasing function for large $\chi$, solving the fine-tuning problem of cosmological constant.
\bibitem{spin triplet}
Two-photon states have either spin triplet symmetric state
or spin singlet anti-symmetric state under exchange of two photons.
Orbital angular momentum part of wave function is classified
by decomposing two plane wave states:
$ e^{i\vec{k}\cdot(\vec{x}_1 - \vec{x}_2)}$
which has all angular momentum combinations starting from S-wave, P-wave $\cdots$.
The lowest symmetric orbital angular momentum states
in expansion of relative coordinate $\vec{x}_1 - \vec{x}_2$ is S-wave.
When this S-wave state is combined with spin triplet,
one has the desired triplet for non-trivial topology.
\end{thebibliography}
\end{document}